\documentclass{ieeeaccess}
\usepackage{cite}
\usepackage{amsmath,amssymb,amsfonts}
\usepackage{algorithmic}
\usepackage{graphicx}
\usepackage{textcomp}
\usepackage{subfigure}
\usepackage{multirow}
\usepackage{algorithm}

\def\BibTeX{{\rm B\kern-.05em{\sc i\kern-.025em b}\kern-.08em
    T\kern-.1667em\lower.7ex\hbox{E}\kern-.125emX}}
\begin{document}
\history{Date of publication xxxx 00, 0000, date of current version xxxx 00, 0000.}
\doi{10.1109/ACCESS.2023.0322000}

\title{SUPPLY: Sustainable multi-UAV Performance-aware Placement Algorithm for Flying Networks}
\author{\uppercase{Pedro Ribeiro},
\uppercase{André Coelho}, \IEEEmembership{Member, IEEE}, and \uppercase{Rui Campos},
\IEEEmembership{Senior Member, IEEE}}

\address{INESC TEC and Faculdade de Engenharia, Universidade do Porto, 4200-465 Porto, Portugal}

\markboth
{Pedro Ribeiro \headeretal: SUPPLY: Sustainable multi-UAV Performance-aware Placement Algorithm for Flying Networks}
{Pedro Ribeiro \headeretal: SUPPLY: Sustainable multi-UAV Performance-aware Placement Algorithm for Flying Networks}

\corresp{Corresponding author: Pedro Ribeiro (e-mail: pedro.m.ribeiro@inesctec.pt)}

\begin{abstract}
Unmanned Aerial Vehicles (UAVs) are used for a wide range of applications. Due to characteristics such as the ability to hover and carry cargo on-board, rotary-wing UAVs have been considered suitable platforms for carrying communications nodes, including Wi-Fi Access Points and cellular Base Stations. This gave rise to the concept of Flying Networks (FNs), now making part of the so-called Non-Terrestrial Networks (NTNs) defined in 3GPP. In scenarios where the deployment of terrestrial networks is not feasible, the use of FNs has emerged as a solution to provide wireless connectivity. However, the management of the communications resources in FNs imposes significant challenges, especially regarding the positioning of the UAVs so that the Quality of Service (QoS) offered to the Ground Users (GUs) and devices is maximized. Moreover, unlike terrestrial networks that are directly connected to the power grid, UAVs typically rely on on-board batteries that need to be recharged. In order to maximize the UAVs’ flying time, the energy consumed by the UAVs needs to be minimized. When it comes to multi-UAV placement, most state-of-the-art solutions focus on maximizing the coverage area and assume that the UAVs keep hovering in a fixed position while serving GUs. Also, they do not address the energy-aware multi-UAV placement problem in networking scenarios where the GUs may have different QoS requirements and may not be uniformly distributed across the area of interest.
In this work, we propose the Sustainable multi-UAV Performance-aware Placement (SUPPLY) algorithm. SUPPLY defines the energy and performance-aware positioning of multiple UAVs in an FN. To accomplish this, SUPPLY defines trajectories that minimize UAVs' energy consumption, while ensuring the targeted QoS levels. The obtained results show up to 25\% energy consumption reduction with minimal impact on throughput and delay.
\end{abstract}


\titlepgskip=-21pt

\maketitle

\section{Introduction}\label{sec:Introduction}
In the past few years, there has been an increasing usage of Unmanned Aerial Vehicles (UAVs) for a wide range of applications. Due to their ability to hover, UAVs are capable of operating almost everywhere. Additionally, they can carry cargo on-board, which makes them suitable platforms for transporting communications nodes, including Wi-Fi Access Points and Cellular Base Stations. This has paved the way for the use of Flying Networks (FNs), composed of UAVs, in order to establish and reinforce wireless coverage and network capacity during temporary events. These temporary events include emergency management scenarios, such as wildfires and floods, and crowded events, such as parades and music festivals. When compared with terrestrial networks, FNs impose additional challenges for optimizing the Quality of Service (QoS) offered to Ground Users (GUs). Since UAVs are not permanently connected to the power grid, they rely on on-board power sources, such as electric batteries, for both communications and movement. Since the energy supplied by the on-board power sources is limited, UAVs need to land frequently for recharging.

The state-of-the-art solutions for UAV placement in FNs are focused on optimizing the QoS offered by means of Flying Access Points (FAPs), typically neglecting the energy consumed for the propulsion of the UAVs; this limits the time the FN is up and running. Recent research works concluded that UAVs are more energy efficient when moving at an optimal speed rather than hovering in a fixed position \cite{8663615}. Based on this conclusion, novel algorithms for defining the UAV trajectory that minimizes the energy consumption without compromising the performance of the FN have been proposed \cite{9759391, Rodrigues2020}. However, the proposed algorithms are focused on a single UAV. When it comes to multiple UAV placement, many state-of-the-art algorithms aim at maximizing the coverage area but disregard that the GUs may have different QoS requirements and may not be uniformly distributed in the area of interest. In order to ensure energy-aware UAV placement, different approaches have been proposed: 1) to minimize the number of UAVs needed to assure the targeted QoS, while positioning the UAVs in hovering state \cite{9764562}; 2) to determine the energy required to assure that the UAVs reach a final destination \cite{9952663}; 3) to optimize UAV altitude, while positioning the UAVs in hovering state \cite{9170560}; 4) to optimize the three-Dimensional (3D) UAV movement \cite{s22051919}. However, these approaches typically do not allow ensuring heterogeneous QoS requirements for GUs. 

To the best of our knowledge, the problem of minimizing the energy consumption for multiple UAVs while meeting target QoS levels associated with different GUs has not been addressed so far. In order to address this problem, we propose the SUPPLY algorithm. SUPPLY defines minimum Signal-to-Noise Ratio (SNR) values in the wireless links between UAVs acting as FAPs and GUs that allow meeting targeted heterogeneous QoS requirements. It then groups the GUs minimizing the number of FAPs needed and defines trajectories for the FAPs that minimize their propulsion energy consumption, while ensuring the targeted QoS levels for the GUs.
SUPPLY is evaluated concerning the FAPs' energy consumption per hour and the FNs' network performance.

The main contributions of this article are threefold:
\begin{itemize}
    \item \textbf{The Sustainable multi-UAV Performance-aware Placement (SUPPLY) algorithm}, which allows defining the energy-aware positioning for multiple UAVs in an FN, while ensuring the targeted QoS levels;
    \item \textbf{The evaluation of the SUPPLY algorithm} in terms of the energy-efficiency gains and the network performance offered in multiple networking scenarios;
    \item \textbf{The Multi-UAV Energy Consumption (MUAVE) simulator}, developed to compute and evaluate the energy consumption of multiple UAVs in an FN.
\end{itemize}

The article is structured as follows. Sections \ref{sec:Background} and \ref{sec:Related Work} present the background and the related work, respectively.
Section \ref{sec:System Model and Problem Formulation} describes the system model and formulates the problem.
Section \ref{sec:SUPPLY} presents the SUPPLY algorithm.
Section \ref{sec:SUPPLY Evaluation} addresses the evaluation of the SUPPLY algorithm regarding UAV energy consumption and network performance.
Section \ref{sec:Discussion} discusses the results obtained for multiple scenarios where the SUPPLY algorithm can be used.
Section \ref{sec:Conclusion} presents the main conclusions and directions for future work.

\section{Background}\label{sec:Background}
In networking scenarios where the deployment of traditional networks is not feasible or limited, alternative solutions for providing wireless connectivity have to be considered. The use of FNs is a promising solution that has emerged in the literature. The advances in UAV design associated with electronics miniaturization have allowed UAVs to be more efficient, lighter, and cheaper. For this reason, the use of FNs has gained increasing interest in the scientific and industrial communities. The main advantage of FNs is their ability for fast deployment almost everywhere, allowing for on-demand wireless communications.
Due to the potential absence of obstacles above the ground, the wireless links between the UAVs that compose the FNs benefit from a strong Line-of-Sight (LoS) component. This allows the use of simplified wireless channel models for network planning, based on the Free-Space Path Loss (FSPL) model. Moreover, in FNs the positions of the UAVs can be fully controlled and adjusted in 3D space. This paves the way for optimizing the positions of UAVs with the ultimate goal of reducing energy consumption while ensuring target QoS levels.

Traditional terrestrial network components are directly connected to the power grid. UAVs rely on on-board power sources, typically electrical batteries. For this reason, the endurance of UAVs is limited by their batteries' capacity. In practice, this means UAVs need to be replaced while their batteries recharge, in order to provide a continuous communications service to GUs.
In an FN, UAVs require energy for two main tasks: 1) communications, including signal processing and transmission; and 2) propulsion, for keeping the UAVs flying, either while moving or hovering. The latter represents the major component when it comes to energy consumption, since it is in the order of hundreds of watts while the former is in the order of watts. In what follows, we present the main influential factors in UAV energy consumption identified in the literature.

Several state-of-the-art works analyze the factors that affect the UAV energy consumption. In \cite{Demir2019}, the authors investigated the energy-efficient deployment of a rotary-wing UAV, considering both flight dynamics and users' QoS requirements, while taking into account the influence of altitude and payload weight. They concluded that as the altitude increases, the UAV power consumption also increases. As the altitude increases the air density decreases, so it is necessary to increase the induced air velocity in order to keep enough thrust to hover. Additionally, the authors of \cite{Demir2019} concluded that an increase in payload also increases the UAV power consumption.

In \cite{8663615} the relation between the propulsion power consumption and the UAV flying speed is described. The authors concluded that UAVs are more energy efficient when they are moving at an optimal constant speed rather than in hovering state. In addition, they proposed a power consumption model for rotary-wing UAVs; this model has been widely used by the scientific community in related works, including \cite{9759391}, \cite{9952663}, and \cite{s22051919}.

In \cite{Muli2022}, a comparative overview of different energy consumption models for UAVs is made. The authors also proposed a data-driven model that was designed to fit a realistic data set \cite{Rodrigues2021}. They concluded that the model created and trained by them is able to fit the real power consumption more accurately than the counterpart models considered. By varying UAV speed, altitude, and payload weight, they were able to conclude that payload weight is the characteristic that mostly influences the energy consumption, followed by the speed and the altitude of the UAV.

More recently, two recent UAV energy consumption models were presented in \cite{9461176} and \cite{9495369}. These models extend the model presented in \cite{8663615} by considering UAV acceleration, which is identified as a relevant factor for the UAV energy consumption. The authors of \cite{9495369} also validated the model proposed in \cite{8663615} for straight line UAV trajectory. Furthermore, they validated their model for Circular trajectories using real flight data. This is the energy consumption model we have considered in this work. We refer to it in more detail in Section \ref{sec:System Model and Problem Formulation} as part of the system model and problem formulation.

\section{Related Work}\label{sec:Related Work}
When it comes to the UAV placement problem, it is possible to group state-of-the-art solutions into single-UAV placement and multi-UAV placement. We refer to each group in the following subsections.

\subsection{Single-UAV Placement}
In \cite{Demir2019}, the authors addressed the energy-aware positioning for a single UAV acting as a Flying Access Point (FAP). The developed algorithm aims at defining the position of the UAV that maximizes the number of users covered, while considering the power constraints and the QoS requirements of the users. The UAV is set to hover at the optimal position that maximizes the coverage area. However, positioning the UAV in a hovering state neglects the fact that UAVs are more energy efficient while moving at an optimal speed.

The placement of a single UAV acting as a gateway was also considered in \cite{RodriguesThesis}, where the author discussed state-of-the-art solutions for gateway positioning in FNs. 
In \cite{9448966}, the Gateway (GW) UAV Placement (GWP) algorithm has been proposed. The GWP algorithm defines the maximum distance that assures the targeted Signal-to-Noise Ratio (SNR) for the wireless links between each FAP and the GW UAV, in order to induce the use of the minimum Modulation and Coding Scheme (MCS) index with a data rate able to accommodate the offered traffic. In 3D space, this results in a sphere for each FAP with a radius equal to the maximum distance allowed. The optimal position for the GW UAV is the subspace defined by the intersection of the resulting spheres for all FAPs in the network. The results obtained confirm that by dynamically defining the position of the GW UAV it is possible to improve the FN performance. Despite the gains obtained in network performance, the GWP algorithm does not take into consideration the energy consumption of the UAVs, considering they keep hovering in a fixed position. In \cite{9759391} and \cite{Rodrigues2020}, the Energy-aware Relay Positioning (EREP) algorithm and its successor Energy and Performance Aware Relay Positioning (EPAP) algorithm were presented. These algorithms build upon the GWP algorithm, considering the same approach for defining the subspace to position the GW UAV. They consider the UAV movement as a key point for the optimization of its energy consumption. After defining a fixed horizontal plane within the subspace where the GW UAV should move, EPAP and EREP compute three candidate trajectories for the GW UAV. The trajectory with the greatest length allows minimizing the UAV energy consumption. The authors concluded that when using EREP and EPAP to define the trajectory of the GW UAV, the network's performance is not compromised when compared to placing the GW UAV in a hovering state, while the GW UAV endurance is increased. However, the proposed algorithms are only applicable to a single GW UAV, while the multiple UAVs that act as FAPs are placed in hovering, which is not the most energy-efficient state.

\subsection{Multi-UAV Placement}
In \cite{8377408}, the authors proposed the NetPlan algorithm, which dynamically adjusts the horizontal position of the FAPs and the Wi-Fi cell ranges based on the traffic demand of the users, in order to improve the FN’s aggregate throughput. The NetPlan algorithm uses the Potential Fields (PF) technique to create force fields that influence the position of the FAPs, attracting FAPs to areas with high traffic demand. The algorithm leads to improvements in QoS metrics, especially in terms of mean throughput. However, NetPlan does not consider the UAVs energy efficiency.

The authors of \cite{9764562} proposed SLICER, an algorithm that enables the placement and allocation of communication resources in slicing-aware flying networks. The SLICER algorithm enables the computation of optimal 3D positions for a minimum number of FAPs, based on GUs positions and their QoS requirements. The algorithm ensures that network slices with target QoS levels are made available as required, while minimizing the utilization of communication resources. However, despite the fact that the FAPs are positioned in suitable positions to meet the QoS levels demanded by the GUs, they are placed in hovering, since the energy-aware UAV placement is not addressed.

In \cite{9170560}, the authors proposed a 3D placement algorithm for multiple FAPs, which are set to hover in a fixed position. Both vertical and horizontal positioning components are addressed, in order to maximize the Global Energy Efficiency (GEE) -- the ratio between the total number of successfully transmitted bits by all User Equipment (UE) and the total energy consumption. After defining the vertical positioning for a single FAP, the horizontal positioning for multiple FAPs is defined by maximizing the total coverage area. Despite defining the altitude at which the UAVs should be placed in order to improve energy efficiency (as concluded in \cite{Demir2019}, energy consumption increases for higher altitudes), the authors consider that the UAVs should be hovering in a fixed position. However, UAVs are more energy efficient when they are moving at an optimal speed.

In \cite{9952663}, an energy-efficient approach for multi-UAV placement was proposed. The goal was to maximize the UAV’s service time and downlink throughput by using a deep reinforcement learning method called Deep-Deterministic Policy Gradient (DDPG), considering that the UAVs, after leaving the first charging station, assure wireless connectivity service to GUs and reach the other charging station without running out of energy. However, the energy efficiency problem was not addressed, since the energy used by the UAVs to reach a charging station was not minimized. Moreover, the movement of the UAVs is not energy-aware, since they can be set to hover over users and travel at non-optimal speeds.

\begin{figure*}[]
\begin{equation}\label{eq:2}
\begin{aligned}
    E(q(t))= & \underbrace{\int_{0}^{T}{P}_{b}\left(1+\frac{3{||v(t)||}^{2}}{{U}_{tip}^2}\right)dt}_{blade\:profile}+\underbrace{\int_{0}^{T}P_{ind}\sqrt{1+\frac{{a_{c}}^{2}(t)}{g^{2}}}\left( \sqrt{1+\frac{{a_{c}}^{2}(t)}{g^{2}}+\frac{{||v(t)||}^{4}}{4{v}_{0}^{4}}}-\frac{{||v(t)||}^{2}}{2{v}_{0}^{2}} \right)^{1/2}dt}_{induced} \\
    & +\underbrace{\int_{0}^{T}\frac{1}{2}{d}_{0}\rho sA{||v(t)||}^{3}dt}_{parasite}+\Delta_{K}
\end{aligned}
\end{equation}
\end{figure*}

In \cite{s22051919}, the authors proposed an energy-efficient UAV movement control model for fair communications coverage. The model divides an area to be covered into different cells, where the center of each cell is referred to as the point of interest with respect to a user. The UAVs should move in order to maximize coverage, by increasing the number of cells covered, and fairness, by homogeneously covering all the cells, while minimizing energy consumption. The authors proposed a novel algorithm, called State-Based Game with Actor-Critic (SBG-AC), to control multiple UAVs in the network. However, regarding UAV placement, the proposed algorithm aims at maximizing the cells covered, considering that they have the same size and are designed to serve a user in their center. The model does not consider a non-uniform positioning of the users and heterogeneous QoS requirements.

In \cite{9994654}, the authors present a comprehensive survey on UAV-assisted wireless networks. The paper provides an overview of solutions for several problems in multiple networking architectures and technologies involving UAVs. Moreover, it points out the challenges imposed by FNs, including UAV energy efficiency. However, concerning multi-UAV placement, there are no solutions addressing energy-efficient UAV movements, while assuring heterogeneous QoS requirements from the GUs.

\section{System Model and Problem Formulation} \label{sec:System Model and Problem Formulation}
\begin{figure*}[]
    \centering
    \includegraphics[width=\textwidth]{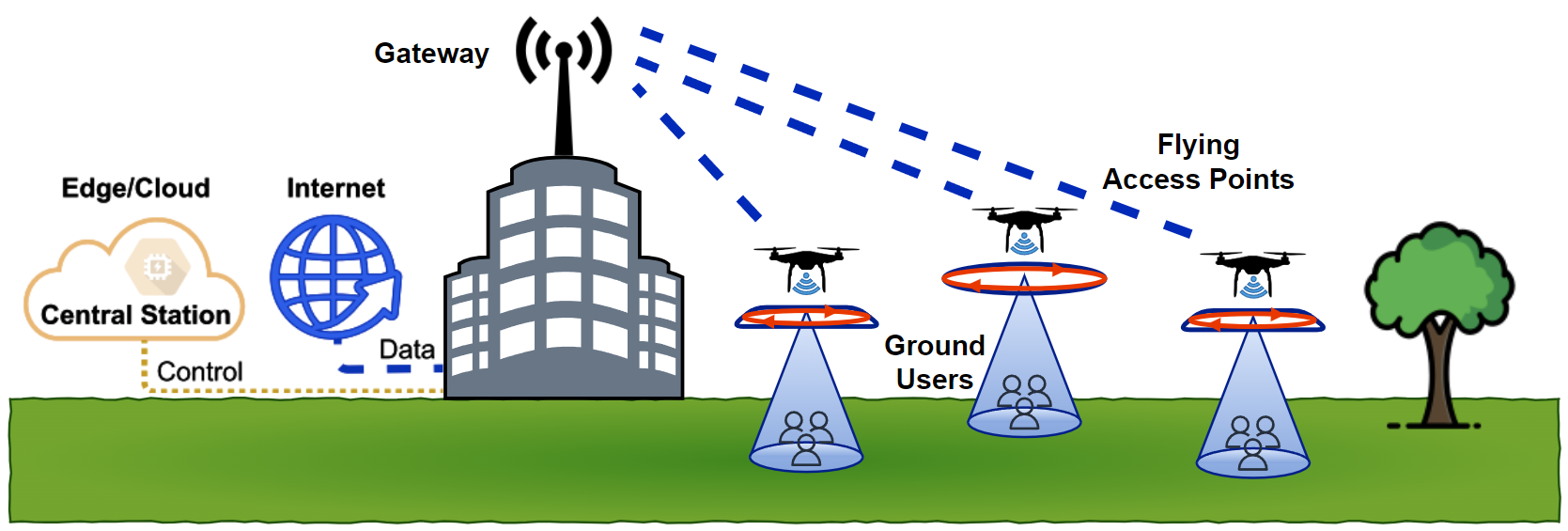}
    \caption{Flying network composed of multiple FAPs following example trajectories, which provide wireless connectivity to multiple GUs, and a gateway that forwards the data traffic to/from the Internet. A remote Central Station monitors the state of the network and defines the placement of the UAVs.}
    \label{fig:Rede}
\end{figure*}

We consider the network architecture depicted in Fig. \ref{fig:Rede}. The network is composed of a Central Station (CS), multiple FAPs and multiple GUs.

\begin{itemize}
    \item \textbf{Central Station (CS)}: responsible for monitoring the state of the network (GUs positions and offered load) and controlling the placement of the UAVs.  
    \item \textbf{Flying Access Points (FAPs)}: rotary-wing UAVs acting as Access Points that provide a radio access network to GUs. The FAPs receive up-to-date placement information from the CS and reposition themselves accordingly.
    \item \textbf{Ground Users (GUs)}: users on the ground that seek wireless connectivity. GUs can have heterogeneous QoS requirements.
\end{itemize}

The problem addressed by this paper lies in defining energy and performance-aware trajectories for the multiple FAPs.
In order to offer different QoS requirements, high enough capacity must be ensured by the wireless links established between the GUs and the FAPs. We assume that the maximum channel capacity is equal to the data rate associated with the MCS index selected by the Medium Access Control (MAC) automatic rate mechanism. In order to achieve higher channel capacity in the wireless links, the value of SNR also needs to be higher. SNR represents the ratio between the power of the received signal (\({P}_{r}\)) over the power of background noise (\({N}_{0}\)), as given by \eqref{eq:3}. The SNR value can be increased in three ways: increasing the transmission power of the FAPs, decreasing the power of background noise, and defining a suitable distance between the communications nodes. Typically, the background noise is not controllable in a real-world environment; we consider it constant for computing the $SNR$ using \eqref{eq:3}.

\begin{equation}\label{eq:3}
    SNR =\frac{{P}_{r}}{{N}_{0}}
\end{equation}

The LoS component is assumed to be strong between the communications nodes, especially due to the potential lack of obstacles induced by the UAV altitude. As such, the FSPL model \cite{goldsmith_2005} is suitable to calculate the power received, as defined in \eqref{eq:4}.

\begin{equation}\label{eq:4}
    \frac{{P}_{r}}{{P}_{t}}= \left[\frac{\lambda}{4\pi d}  \right]^2
\end{equation}

In \eqref{eq:4}, \({P}_{r}\) is the received power, \({P}_{t}\) is the transmission power, \(d\) is the Euclidean distance between the transmitter and the receiver nodes, and \(\lambda\) is the wavelength.
The received power falls off in inverse proportion to the square of the distance \(d\) between the transmitter and the receiver. Considering a constant transmission power \({P}_{t}\), a target received power \({P}_{r}\) can be ensured by defining a suitable distance \(d\).

We assume the wireless medium is shared and every FAP and GU can listen to each other. The fair share is determined by dividing the maximum capacity of the wireless link established between each GU and the serving FAP. The maximum capacity is assumed to be equal to the data rate associated with the used MCS index divided by the number of GUs sharing the medium.
Considering these conditions, we use the Carrier Sense Multiple Access with Collision Avoidance (CSMA/CA) mechanism for MAC.

As referred to in Section \ref{sec:Background}, UAVs are more energy efficient while moving at an optimal speed. The optimal speed depends on the type of trajectory that the UAV is following.
In order to calculate the FAPs energy consumption, we use the state-of-the-art energy consumption model defined in \eqref{eq:2}, which is presented in \cite{9495369}.

In \eqref{eq:2}, \({P}_{b}\) and \({P}_{ind}\) are two constants representing the \textit{blade profile power} and \textit{induced power}. In hovering state, \({U}_{tip}\) is the tip speed of the rotor blade, \({v}_{0}\) is the mean rotor induced velocity, \({d}_{0}\) and \(s\) are respectively the fuselage drag ratio and rotor solidity, \({\rho}\) stands for the air density, and \(A\) denotes the rotor disc area. \({v(t)}\) and \({{a_{c}}^{2}(t)}\) are respectively the flying speed and the centrifugal acceleration at the time instant $t$.
For the Circular movement, it is possible to model the UAV power consumption using \eqref{eq:5} derived from the model given by \eqref{eq:2}, considering ${a}_{c}=\frac{{V}^{2}}{r}$ and $||v(t)||=V$, where $r$ denotes the Circular radius and $V$ the UAV flying speed.

\begin{equation}\label{eq:5}
\begin{split}
P(V,r) = & \underbrace{{P}_{b}\left(1+\frac{3{V}^{2}}{{U}_{tip}^2}\right)}_{blade\:profile} \\
 & + \underbrace{P_{ind}\sqrt{1+\frac{{V}^{4}}{r^{2}g^{2}}}\left( \sqrt{1+\frac{{V}^{4}}{r^{2}g^{2}}+\frac{{V}^{4}}{4{v}_{0}^{4}}}-\frac{{V}^{2}}{2{v}_{0}^{2}} \right)^{1/2}}_{induced} \\
 & + \underbrace{\frac{1}{2}{d}_{0}\rho sA{V}^{3}}_{parasite}
\end{split}
\end{equation}

\begin{figure}[]
    \centering
    \includegraphics[width=0.45\textwidth]{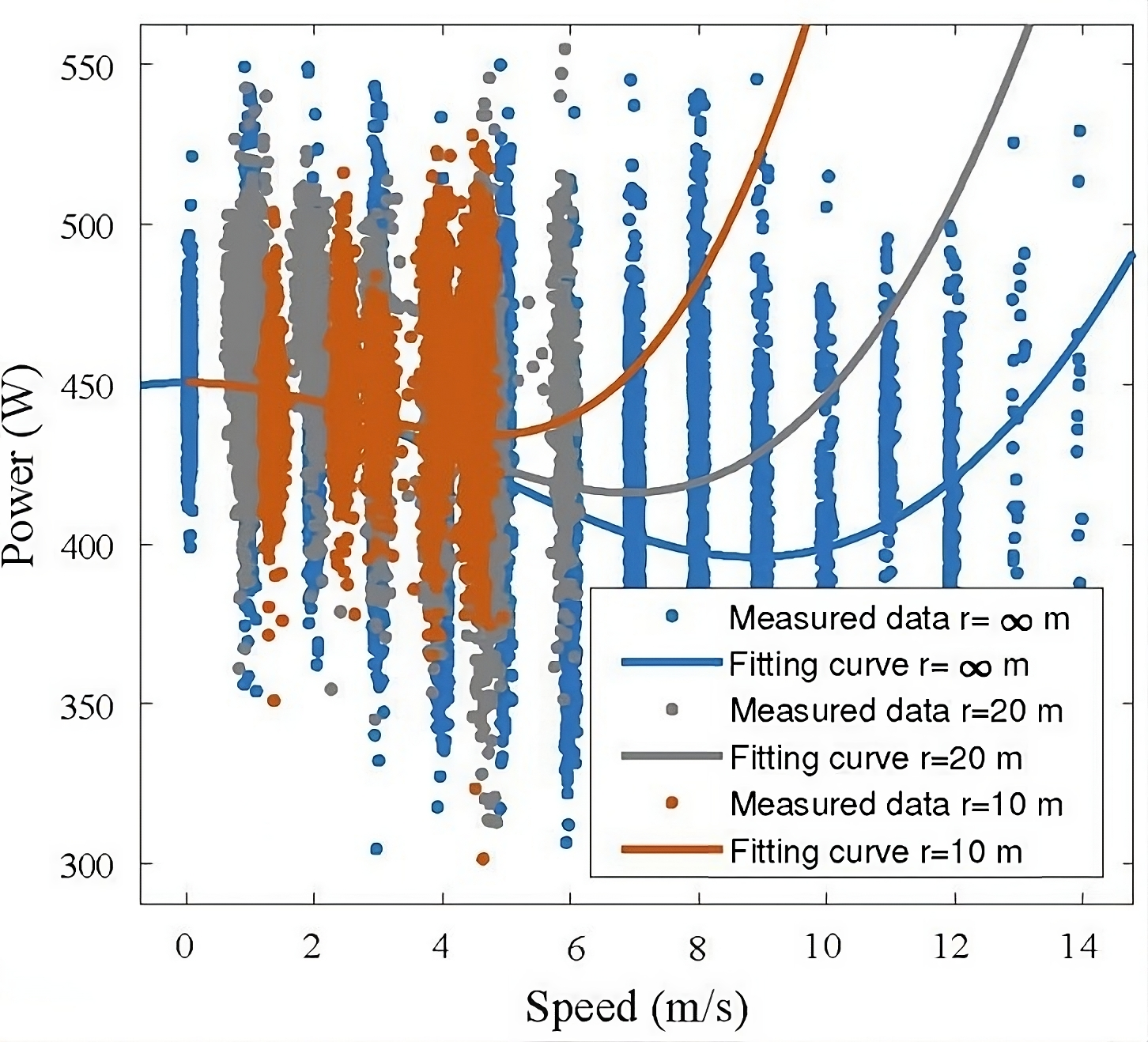}
    \caption{Fitting curves of power consumption in Circular flight for a rotary-wing UAV \cite{9495369}.}
    \label{fig:speed_power_radius}
\end{figure}

As given by \eqref{eq:5}, the power consumption depends on the flying speed and the Circular radius. For each given Circular radius, there is an optimal speed that leads to the minimum power consumption. The plot of Fig. \ref{fig:speed_power_radius} shows that the power consumption increases and the optimal speed decreases with the decrease of the radius of the trajectory of the UAV. For straight line movement (where the radius approaches infinity) the optimal speed has the highest value, and the power consumption reaches its lowest value. Based on this, it is possible to conclude that the FAPs moving in a straight line at the optimal speed is the deployment approach that allows minimizing UAV energy consumption. The model defined in \eqref{eq:5} allows the calculation of the optimal speed and the corresponding power consumption for curved-based trajectories with well-defined radii.

For a given period $T$ where the FN is active, with $0\leq t\leq T$, the FN is modeled as follows. $G=\left\{{GU}_{1},..., {GU}_{N}\right\}$ is the set of GUs, where $N$ is the number of GUs, with $GU_i$ positioned at ${Q}_{i} = ({x}_{i}, {y}_{i}, 0)$. $F=\left\{{FAP}_{1},..., {FAP}_{M}\right\}$ is the set of FAPs, where $M$ is the number of FAPs, with $FAP_j$ positioned at ${Q}_{j}(t) = \left[{x}_{j}(t), {y}_{j}(t),{z}_{j}(t)\right]$. $L \subseteq  G \times F$ is the set of links between the GUs and FAPs.

Let us assume that each ${GU}_{i}$, $i \in G$, sends a traffic flow ${T}_{i,j}$ towards ${FAP}_{j}$, $j \in F$. Each ${GU}_{i}$ is served by a single $FAP_j$.
The objective is to minimize the number of FAPs $M$ and determine a trajectory ${Q}_{j}(t) = \left[{x}_{j}(t), {y}_{j}(t),{z}_{j}(t)\right]$ for each ${FAP}_{j}$ so that the energy consumed by all FAPs is minimized and the transfer of each traffic flow ${T}_{i,j}$ with data rate ${B}_{i,j}$, in bit/s, is guaranteed. For that purpose, a wireless link with high enough capacity ${C}_{i,j}(t)$ to accommodate ${B}_{i,j}$ must be guaranteed between each ${GU}_{i}$ and a ${FAP}_{j}$. Moreover, the aggregate capacity of the wireless links must be lower than or equal to the maximum channel capacity ${C}^{MAX}$.

\begin{subequations}
    \begin{equation}\label{eq:6}
        minimize \sum_{j=1}^{M}\int_{0}^{T}P_{j}\left(\left \|Q_{j}(t) \right \|\right ) dt
    \end{equation}
Subject to:
    \begin{equation} \label{b}
        P_{j}(t)\le P_{j}^{MAX}, \forall t\in \left [0,T  \right ]
    \end{equation}
    \begin{equation} \label{c}
        Q_{k}(t)\neq Q_{l}(t), \forall k,l \in F, \forall t\in \left [0,T  \right ]
    \end{equation}
    \begin{equation} \label{d}
        z_{j} \ge 0, \forall j \in F
    \end{equation}
    \begin{equation} \label{e}
        \sum_{j=1}^{M}L_{i,j} = 1, \forall i \in G 
    \end{equation}
    \begin{equation} \label{f}
        0 \le B_{i,j}(t)\le C_{i,j}(t), \forall i \in G, \forall j \in F, \forall t\in \left [0,T  \right ]
    \end{equation}
    \begin{equation} \label{g}
        \sum_{i=1}^{N}{C}_{i,j}(t)\le {C}^{MAX},\forall j \in F, \forall t\in \left [0,T  \right ]
    \end{equation}

\end{subequations}

\begin{itemize}
    \item (\ref{b}) ensures that the propulsion power enabling the operation of any $FAP_{j}$ does not exceed its maximum propulsion power $P_{j}^{MAX}$.
    \item (\ref{c}) assures that the position defined for each FAP is different from the positions of any other FAP. This avoids collisions between FAPs.
    \item (\ref{d}) guarantees that all the FAPs and GUs are positioned at ground level z=0 or above.
    \item (\ref{e}) ensures the existence of one and only one link $L$ between each GU and a FAP.
    \item (\ref{f}) guarantees that the data rate $B_{i,j}(t)$ of the traffic flow $T_{i,j}(t)$ between $GU_{i}$ and $FAP_{j}$ is lower than or equal to the capacity $C_{i,j}(t)$ of the link.
    \item (\ref{g}) ensures that the aggregate capacity of the wireless links between all FAPs and GUs is lower than or equal to the maximum capacity ${C}^{MAX}$ of the wireless channel.
\end{itemize}

\section{Sustainable multi-UAV Performance-aware Placement Algorithm} \label{sec:SUPPLY}
In order to solve the problem formulated in Section \ref{sec:System Model and Problem Formulation}, we propose the Sustainable multi-UAV Performance-aware Placement (SUPPLY) algorithm. Considering networking scenarios where the GUs are located in a given area of interest, SUPPLY groups the GUs to minimize the number of FAPs needed. Furthermore, SUPPLY defines energy-efficient trajectories for the FAPs.
The algorithm builds upon the EREP and SLICER algorithms proposed in \cite{Rodrigues2020} and \cite{9764562} respectively, allowing to:

\begin{itemize}
    \item Minimize the number of FAPs needed to assure the target QoS requirements;
    \item Define trajectories for the FAPs that minimize the energy consumption and do not compromise the target QoS.
\end{itemize}

SUPPLY is divided into two phases.
In the first phase, SUPPLY groups the GUs so that the number of FAPs needed is minimized.
In the second phase, it defines the energy-efficient trajectories for the FAP serving each group of GUs.

\begin{algorithm}[]
\caption{SUPPLY Algorithm}\label{alg:1}
\hspace*{\algorithmicindent} \textbf{Input}: GUs positions, GUs offered load, possible FAP positions\\
\hspace*{\algorithmicindent} \textbf{Output:} Trajectories of the FAPs  
\begin{algorithmic}[1]
\FORALL{Possible FAP positions}\label{l:1}
    \STATE Calculate SNR values for potential links with GUs \label{l:2}
    \STATE Estimate capacity of wireless links GU-FAP\label{l:3}
\ENDFOR
\STATE Groups of GUs $\leftarrow$  Optimizer (GUs positions, possible FAP positions, GUs offered load, capacity of the wireless links GU-FAP)\label{l:4}
\FORALL{Groups of GUs}
    \STATE Define minimum target SNR for each GU \label{l:7}
    \STATE Calculate maximum Euclidean distance that ensures target SNR for each GU\label{l:8}
    \STATE Calculate intersection area \label{l:9}
    \IF{Intersection area overlaps with previous calculated intersection's areas} \label{l:10}
        \STATE Remove overlapped area from intersection area to avoid collisions between FAPs \label{l:11}
    \ENDIF
    \STATE Compute centroid of intersection area \label{l:14}
    \STATE Compute perimeter of intersection area \label{l:15}
    \STATE Compute Circular trajectory \label{l:16}
    \STATE Compute Inner Elliptic trajectory \label{l:17}
    \STATE Compute Elliptic trajectory \label{l:18}
    \STATE Choose most energy-efficient FAP trajectory \label{l:19}
\ENDFOR
\end{algorithmic}
\end{algorithm}

The FAPs are expected to be placed as close to the ground as possible, since it has been demonstrated in the literature that higher altitudes lead to increased power consumption, as explained in Section \ref{sec:Related Work}. Also, placing the FAPs close to the GUs allows improving the SNR offered. We consider a fixed altitude of 6 meters ($z=6$) for the FAPs, in order to avoid collisions with people and obstacles on the ground. Considering a fixed altitude also allows reducing the complexity of the problem by considering a lower number of possible FAP positions. In order to define all possible FAP positions within the specified area, SUPPLY takes into account a step of 1 meter in both the x and y axes. This is crucial to assure that GUs are grouped optimally and that the number of FAPs used is minimized.

The initial step of the algorithm consists in calculating the SNR values for the links between the GUs and all the possible FAP positions (Line \ref{l:2} of Alg. \ref{alg:1}). To calculate the SNR values, we assume a transmission power of 20 dBm for all the nodes. Additionally, we add a 1 dB SNR margin to the links in order to ensure consistent network performance. This value can be fine-tuned using a mechanism that is left for future work.
The calculated SNR values are then used to determine the capacity of the links by correlating them with the minimum data rate provided by each MCS index (Line \ref{l:3} of Alg. \ref{alg:1}).
SUPPLY considers information about the GUs positioning, the possible FAPs positioning, the GUs offered load, and the capacity obtained for all the links established with the FAPs, and provides as output the clusters of GUs associated with each FAP that minimize the number of required FAPs. This is solved as an optimization problem.
In its current version, SUPPLY employs the Gurobi optimizer \cite{Gurobi} (Line \ref{l:4} of Alg. \ref{alg:1}) for this purpose. However, other solvers may be considered. During this step, SUPPLY ensures that each GU can be served by a FAP. The optimal FAP positions and energy-efficient trajectories are calculated in the second phase.

\begin{figure*}[]
    \centering
    \includegraphics[width=0.75\textwidth]{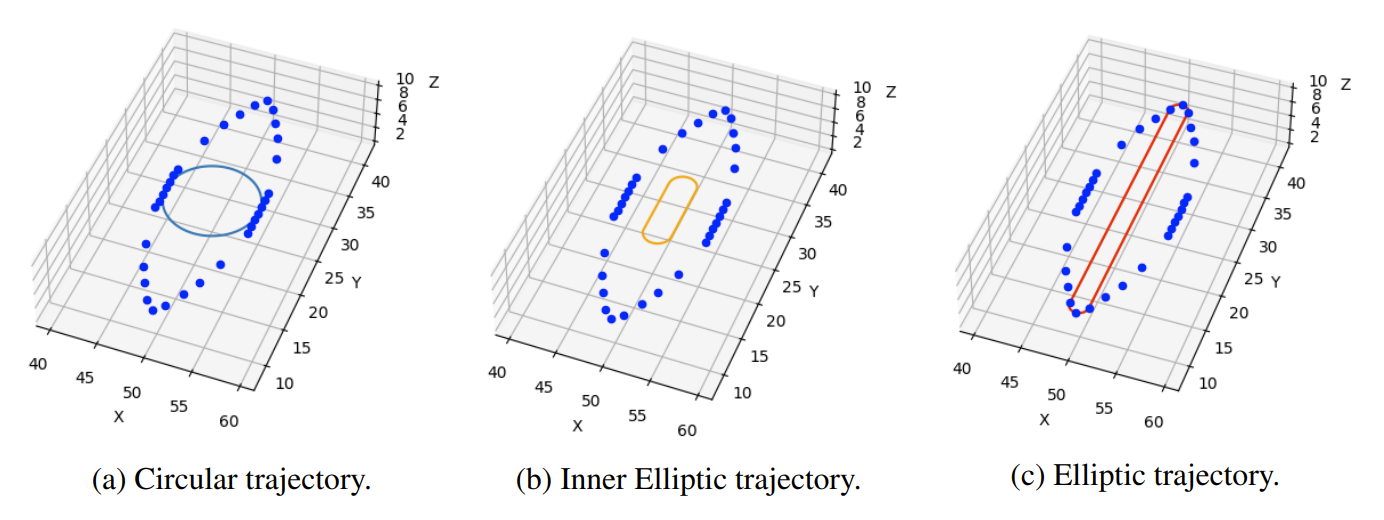}
    \caption{SUPPLY trajectories. The blue points represent the perimeter of the intersection area at a step of 1 meter in the x and y axis. The colored continuous lines within the intersection area represent the SUPPLY trajectories.}
    \label{fig:Trajectories}
\end{figure*}

SUPPLY defines the maximum distance that ensures the targeted SNR for the wireless links established between the GUs and the FAPs (Lines \ref{l:7} and \ref{l:8} of Alg. \ref{alg:1}). In 3D space, this results in a sphere for each GU, with a radius equal to the maximum distance defined.
The optimal position for each FAP is the subspace defined by the intersection of the spheres associated with all GUs served by the FAP, as defined in \cite{9448966}. Considering that altitude variations are not energy efficient, the FAPs should move within the subspace at a fixed altitude.
The altitude is chosen considering the plane of the intersection volume that maximizes the resulting intersection area (Line \ref{l:9} of Alg. \ref{alg:1}); the increased area allows to define more energy-efficient trajectories to be completed by the FAP, thus leading to reduced energy consumption. Since the GUs are on the ground (altitude equal to 0 meters), the resulting intersection areas decrease with the increase in altitude. In SUPPLY, we consider a constant altitude of 6 meters. The algorithm assures that the resulting intersection areas do not overlap, in order to avoid collisions between FAPs. For that purpose, any overlapped areas are removed (Lines \ref{l:10} and \ref{l:11} of Alg. \ref{alg:1}). The optimal position for each FAP is the centroid of the corresponding intersection area (Line \ref{l:14} of Alg. \ref{alg:1}); this is the position considered for the hovering state. Additionally, the perimeter of the intersection area is also delineated, as it is crucial to establish the new trajectories associated with the centroid (Line \ref{l:15} of Alg. \ref{alg:1}).
After determining the intersection areas and their centroid and perimeter, the algorithm defines three possible geometric trajectories: Circular, Inner Elliptic, and Elliptic (Lines \ref{l:16}, \ref{l:17} and \ref{l:18} of Alg. \ref{alg:1}). These trajectories were chosen in order to utilize the available area to minimize the FAPs' energy consumption, considering that FAPs are more energy-efficient while moving at optimal speeds in straight lines or following curved trajectories with large radii. The energy consumed by a FAP to complete each trajectory is computed using the power consumption model given by \eqref{eq:5}. Representative examples for three types of trajectories are depicted in Fig. \ref{fig:Trajectories}.
The first trajectory defined by SUPPLY is the Circular trajectory. The SUPPLY algorithm defines the radius of the Circular trajectory as being the minimum distance from the area's centroid to its perimeter. Then, by connecting the equidistant points around the centroid at the defined radius, a Circular trajectory is achieved. This approach assures that each FAP remains within the defined area while defining a trajectory with the maximum possible radius.
The second trajectory is named Inner Elliptic, as its shape closely resembles an ellipse and it is defined within the circumference associated with the Circular trajectory. This trajectory is also centered at the area's centroid and is composed of two straight line segments and two semi-circles, with a radius of 30\% of the radius used to define the Circular trajectory. The radius for the semi-circles is chosen to keep the trajectory's size proportional to the size of the limiting area. The trajectory follows the orientation of a straight line segment connecting the farthest apart perimeter points. For the cases where the area is not Circular, this implies that the FAP does not reach the perimeter. This approach has the potential to enhance network performance, since the links established between the GUs and the FAP will have a higher SNR value, allowing for higher data rate values.
The last trajectory is referred to as Elliptic. Unlike the Inner Elliptic, this trajectory is not confined to the circumference of the Circular trajectory. Instead, it utilizes all the area to maximize the straight line movement. As with the Inner Elliptic trajectory, the Elliptic trajectory is also composed of two straight line segments and two semi-circles, while following the orientation of the straight line segment connecting the farthest apart perimeter points. However, it is centered in the middle of this straight line segment instead of the area's centroid. Furthermore, the radius of the semi-circles is given by the smallest distance between the straight line segment and the closest perimeter point, excluding the two used to form the straight line segment. This assures that the FAP is within the area. If the intersection area is too small, not allowing the creation of the trajectories, then the FAP is set to hover in the optimal position. The SUPPLY algorithm then chooses the trajectory that leads to better energy efficiency gains: the trajectory with the least energy consumption (Line \ref{l:19} of Alg. \ref{alg:1}).

\section{SUPPLY Evaluation} \label{sec:SUPPLY Evaluation}
The SUPPLY algorithm was evaluated in terms of energy consumption and network performance. Based on the system model presented in Section \ref{sec:System Model and Problem Formulation}, we defined multiple scenarios with the objective of evaluating how SUPPLY is able to define energy-efficient trajectories while assuring the targeted QoS requirements. All the scenarios consisted of a variable number of GUs, randomly distributed in an area of 100 $\times$ 100 meters, with heterogeneous QoS requirements.

We considered a single channel for each group, since the focus of this work is not the management of the wireless channel allocation. The FAPs act as independent Access Points.
For traffic generation purposes, we considered a maximum channel capacity of 500 Mbit/s, which is divided by the number of GUs sharing the channel. Increasing the number of GUs sharing the channel leads to lower capacity available for the wireless link established with each user. In order to ensure that the channel was not saturated, which would impact the performance results, the offered load was reduced when increasing the number of GUs.

For the network performance evaluation, two QoS metrics were considered:
1) average throughput -- the average number of bits received per second by a FAP, considering all GUs in the group;
2) average packet delay -- the average time a packet takes to reach the sink application of each FAP, starting from the moment the packet is generated at the GU. The packet delay includes queuing, transmission, and propagation delays. Since each FAP may receive packets from multiple GUs, an average of the delay associated with those packets was calculated at each second during the simulation.
In the performance evaluation carried out, a special emphasis was given to throughput, as it allows to prove the traffic-awareness of the SUPPLY algorithm, considering as reference the data rate associated with the generated traffic flows.

\subsection{Simulation Setup}
To perform the energy consumption evaluation when using SUPPLY, we employed the MUAVE simulator \cite{MUAVE}, a Python-based simulator that allows considering multiple GUs groups and their data rate requirements for computing the energy consumption for multiple FAPs providing wireless connectivity to GUs. MUAVE builds upon the UAV Power Simulator proposed in \cite{UPS} for computing the power consumption for a single UAV acting as a gateway.

In order to obtain the intersection areas, the SUPPLY algorithm considers the minimum SNR threshold for each MCS index as well as the achievable data rates. The values considered in the simulator are presented in Table \ref{tab:SNRvsDataRate}. These values were obtained by running an ns-3 simulation with 2 nodes. While progressively increasing the distances between the 2 nodes to reduce the SNR, we registered the MCS index selected by the \emph{IdealWifiManager} MAC automatic rate for each distance. For those distances, we calculated the SNR values achieved for the wireless link established, using the FSPL model and a constant noise power.

\begin{table}
\centering
\caption{Relation between minimum target SNR and expected Data Rate for a number of ground users $N$, considering the IEEE 802.11ac technology and 160 MHz channel bandwidth.}
\label{tab:SNRvsDataRate}
\setlength{\tabcolsep}{3pt}
\begin{tabular}{|c|c|}
\hline
\bfseries SNR (dB) & \bfseries Data Rate (Mbit/s)  \\
\hline
\hline
13.1  & 53/$N$   \\ 
13.6  & 103/$N$  \\ 
16.1  & 152/$N$  \\ 
19.5  & 198/$N$  \\ 
22.6  & 287/$N$  \\ 
27.1  & 368/$N$  \\ 
28.4  & 405/$N$  \\ 
29.9  & 447/$N$  \\ 
34.1  & 518/$N$  \\ 
35.3  & 553/$N$  \\
\hline
\end{tabular}
\end{table}

Regarding UAV energy consumption, we considered several parameters that allow characterizing the UAV model and the environment. We employed the values presented in \cite{8663615}.  These parameters include: UAV weight      $W$ = 20 $N$, rotor radius $R$ = 0.4 $m$, blade angular velocity $\Omega$ = 300 $rad/s$, incremental correction factor to induced power $k$ = 0.1, profile drag coefficient $\delta$ = 0.012, air density $\rho$ = 1.225 $kg/m^3$,
rotor disc area $A = \pi R^2$ = 0.503 $m^2$, tip speed of the rotor blade $U_{tip}$ = 120 $m/s$, fuselage drag ratio $d_{0}$ = 0.6, mean rotor induced velocity in hovering state $v_{0}=\sqrt{\frac{W}{2\rho A}}$ = 4.03, rotor solidity $s$ = 0.05, blade profile power in hovering state $P_{b} = \frac{\delta}{8}\rho sA\Omega^{3}R^{3}$ = 79.86 and induced power in hovering state $P_{ind} = (1+k)\frac{W^{3/2}}{\sqrt{2\rho A}}$ = 88.63.
Additionally, since we rely on an extended version of the energy model proposed in \cite{8663615}, we also considered the gravitational acceleration $g$ = 9.8 $m/s^{2}$.

In order to evaluate the network performance, we used the ns-3 simulator \cite{ns3}. For each node representing a GU, a Network Interface Card (NIC) was configured as a Station (STA) in Infrastructure mode. The IEEE 802.11ac standard was considered and the two available 160 MHz channels, 50 and 114, were used. UDP traffic was employed, with the packets being generated according to a Poisson distribution. Each data packet had a size of 1400 bytes. To manage the physical data rates of the wireless links, the MAC automatic rate mechanism \emph{IdealWifiManager} was used.
The simulation duration was set to 70 seconds, with an additional 30 seconds to start the applications and allow the network to reach a steady state. To manage the queues and mitigate the bufferbloat problem in the simulation, the Controlled Delay (CoDeL) algorithm \cite{CoDeL} was employed. For channel modeling, we used the FSPL model. To enhance the realism of the validation process, we also introduced a Rician fast-fading component, considering a Rician K-factor equal to 13 dB, which was derived in \cite{ALMEIDA2021102525} based on experimental results. The values of the simulation parameters are summarized in Table \ref{tab:SimParameters}.

\begin{table}
\caption{\textbf{Simulation Parameters.}}
\label{tab:SimParameters}
\centering
\setlength{\tabcolsep}{3pt}
\begin{tabular}{|p{175 pt}|c|}
\hline
\bfseries Parameter & \bfseries Value  \\ \hline \hline
UAV weight ($W$) & 20 $N$           \\
Rotor radius ($R$) & 0.4 $m$         \\
Blade angular velocity ($\Omega$) & 300 $rad/s$      \\
Incremental correction factor to induced power ($k$) & 0.1   \\
Profile drag coefficient ($\delta$) & 0.012            \\
Air density ($\rho$) & 1.225 $kg/m^3$     \\
Gravitational acceleration ($g$) & 9.8 $m/s^{2}$    \\
Rotor disc area ($A$) & 0.503 $m^2$        \\ 
Tip speed of the rotor blade ($U_{tip}$) & 120 $m/s$        \\ 
Fuselage drag ratio ($d_{0}$)  & 0.6              \\
Mean rotor induced velocity in hovering state ($v_{0}$) & 4.03             \\ 
Rotor solidity ($s$) & 0.05             \\ 
Blade profile power in hovering state ($P_{b}$) & 79.86            \\ 
Induced power in hovering state ($P_{ind}$) & 88.63 \\
Wi-Fi standard & IEEE 802.11ac        \\ 
Channel bandwidth & 160 $MHz$       \\ 
Wireless channels & 50, 114        \\
Channel frequency & 5250 $MHz$      \\
Guard interval & 800 $ns$         \\
Transmission power & 20 $dBm$ \\
Noise power & -85 $dBm$ \\
Traffic type & UDP    \\ 
Packet size & 1400 $B$      \\
Simulation time & 70 $s$        \\
Rician K-factor & 13 $dB$ \\
\hline
\end{tabular}
\end{table}

\subsection{Evaluation Under Typical Networking Scenarios}
In this section, we present the results for typical networking scenarios. We considered 2 scenarios with 2 GUs, 2 scenarios with 5 GUs, and 2 scenarios with 10 GUs, which are characterized in Table \ref{tab:NetworkScenarios}. For each scenario, the positions and offered load of the GUs were randomly defined. For each number of GUs, one of the scenarios resulted in the formation of a single group (1 FAP used) and the other led to the formation of 2 groups (2 FAPs used). The results of each scenario are based on all values gathered from 10 simulation runs.
The average throughput results measured at each FAP are presented by means of a Complementary Cumulative Distribution Function (CCDF) while the average packet delay results are represented by a Cumulative Distribution Function (CDF). The CDF $F(x)$ indicates the proportion of collected packets that exhibited delay less than or equal to $x$, while the CCDF $F(x)$ quantifies the percentage of time during which the average throughput was greater than $x$. 

\renewcommand{\arraystretch}{1.5}
\begin{table*}[]
\caption{\textbf{GUs arbitrary positions and GUs offered load for the networking scenarios considered.}}
\label{tab:NetworkScenarios}
\centering
\begin{tabular}{|c|c|c|c|}
\hline
\multicolumn{2}{|c|}{\textbf{Network Scenario}} & \textbf{GUs Positions (x,y,z)}  & \textbf{GUs offered load (Mbit/s)}   \\  \hline \hline
\multirow{2}{*}{2 GUs}  & 1 FAP  & (47,32,0), (52,71,0) &  200, 117             \\ \cline{2-4} 
                        & 2 FAPs & (27,17,0), (91,60,0) & 174, 208              \\ \hline
\multirow{4}{*}{5 GUs} & \multirow{2}{*} {1 FAP} & (19,62,0), (85,46,0), & 36, 27,                     \\ 
                        &                        & (86,53,0), (2,9,0), (52,88,0) & 19, 14, 23          \\ \cline{2-4} 
                        & \multirow{2}{*} {2 FAPs} & (58,56,0), (24,88,0), & 73, 42,                   \\
                        &                          & (35,92,0), (60,25,0), (51,14,0) & 87, 11, 71        \\ \hline
\multirow{6}{*}{10 GUs} & \multirow{3}{*} {1 FAP} & (69,83,0), (68,91,0), (26,16,0), & 9, 6, 1,            \\ 
                        &                       & (67,8,0), (38,21,0), (23,71,0), & 5, 3, 6,             \\
                        &                       & (60,34,0), (8,31,0), (59,59,0), (20,79,0) &  7, 5,
8, 6\\ \cline{2-4} 
                        & \multirow{3}{*} {2 FAPs} & (47,27,0), (29,55,0), (42,30,0), & 38, 32, 29,          \\
                        &                         & (2,17,0), (19,49,0), (68,66,0), & 39, 43, 21,            \\ 
                        &                      & (39,15,0), (79,76,0), (84,7,0), (82,9,0) & 8, 49, 6, 14           \\ \hline
\end{tabular}
\end{table*}

\subsubsection{2 GUs}
\begin{figure}[]
    \centering
    \includegraphics[width=0.45\textwidth]{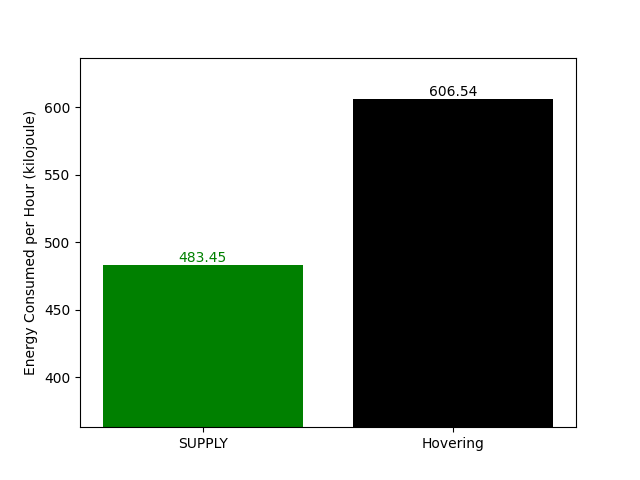}
    \caption{Comparison of energy consumption per hour for 2 GUs in the same group.}
    \label{fig:2G1F_Energy}
\end{figure}

The first scenario considered 2 GUs, each with a randomly selected offered load up to an aggregated value of 500 Mbit/s. The results for energy consumption per hour are presented in Fig. \ref{fig:2G1F_Energy}. SUPPLY led to an energy consumption per hour of 483.45 kJ, while the hovering state resulted in 606.54 kJ. Every trajectory defined by the SUPPLY algorithm reduces energy consumption. Still, the SUPPLY algorithm selected the Circular trajectory, as it leads to less energy consumption, with a reduction of 20\% in relation to the hovering state.

\begin{figure}[]
    \centering
    \includegraphics[width=0.45\textwidth]{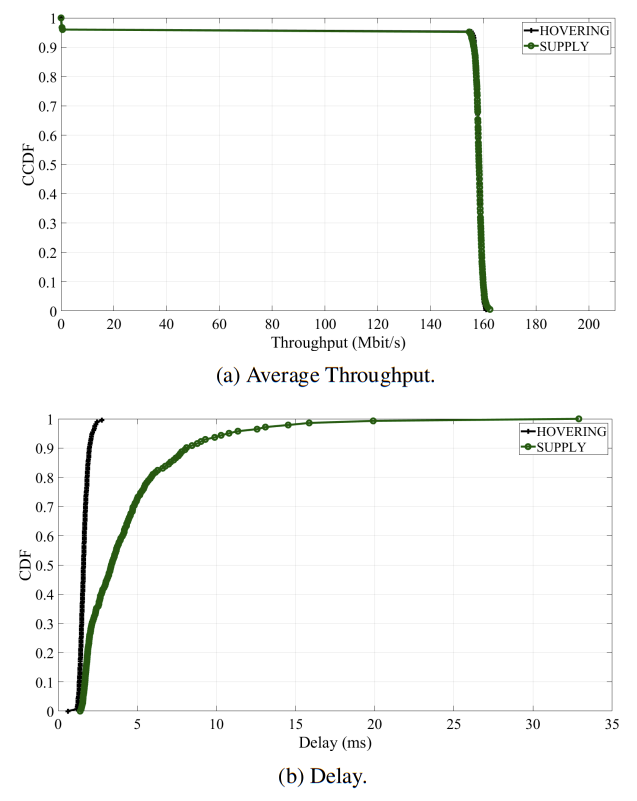}
    \caption{Network performance results for 2 GUs in the same group.}
    \label{fig:2G1F}
\end{figure}

Regarding network performance, the results for the average throughput show that for the trajectory defined by SUPPLY there is no significant impact in relation to the hovering state ($<0.5\%$), assuring the expected average throughput of 158.5 Mbit/s. In terms of delay, there is a small increase when using SUPPLY: approximately a 5 ms increase for the $90^{th}$ percentile, as illustrated in Fig. \ref{fig:2G1F}. The increased delay is justified by the greater distances between the FAP and the GUs that result from following the trajectory. When the FAP is farther away from the GUs, the link capacity is slightly reduced due to the decrease in SNR. For that reason, packets are held in transmission queues for a longer period of time.

\begin{figure}[]
    \centering
    \includegraphics[width=0.45\textwidth]{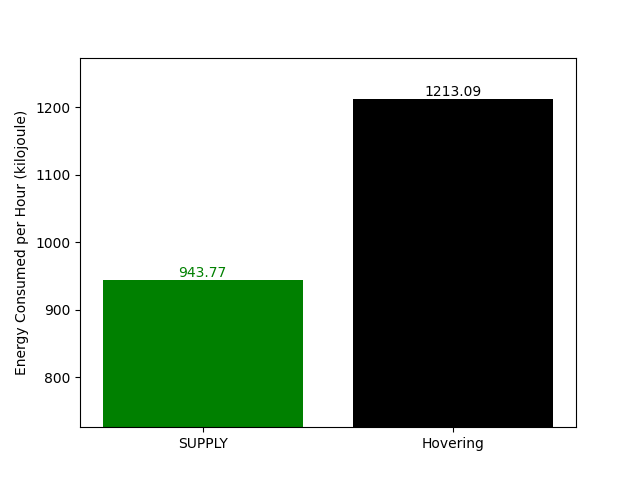}
    \caption{Comparison of energy consumption per hour for 2 GUs in 2 separate groups.}
    \label{fig:2G2F_Energy}
\end{figure}

The second scenario, consisting of 2 GUs, led to the usage of 2 FAPs, each serving one of the GUs. In this scenario, the SUPPLY algorithm chose the Circular trajectory for both FAPs resulting in an energy consumption per hour of 943.77 kJ. This represents a reduction in energy consumption of 22\% when compared with the hovering state (1213.09 kJ), as depicted in Fig. \ref{fig:2G2F_Energy}.

\begin{figure}[]
    \centering
    \includegraphics[width=0.45\textwidth]{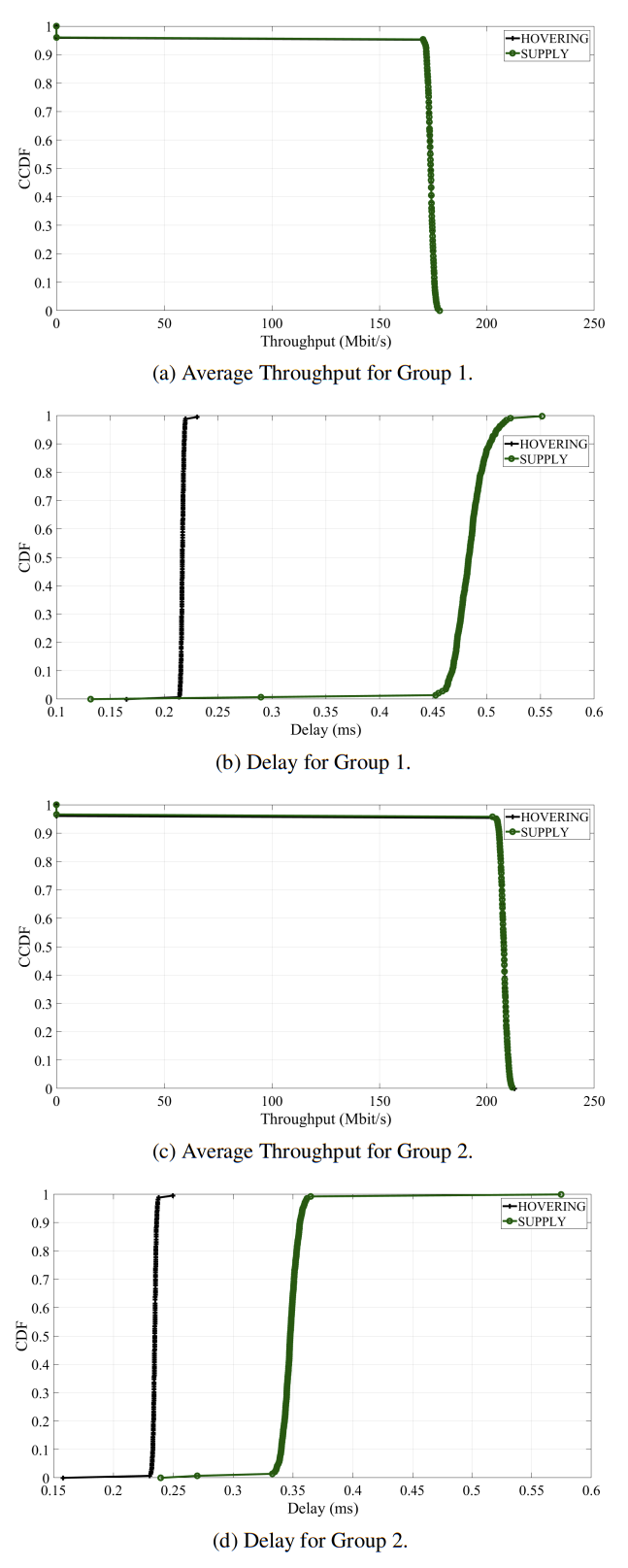}
    \caption{Network performance results for 2 GUs in separate groups.}
    \label{fig:2G2F}
\end{figure}

Regarding network performance, for the average throughput, SUPPLY does not affect the results achieved in relation to the hovering state.
In terms of delay, the percentiles indicate an increase in both GUs. However, the difference is negligible, since the values are below 1 ms in both cases, as depicted in Fig. \ref{fig:2G2F}.

\subsubsection{5 GUs}
\begin{figure}[]
    \centering
    \includegraphics[width=0.45\textwidth]{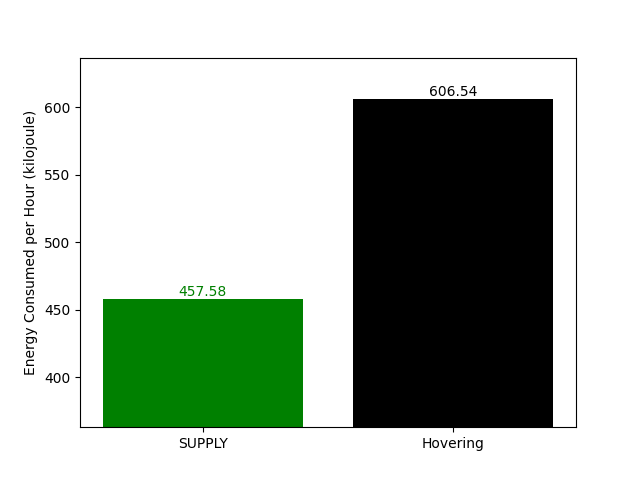}
    \caption{Comparison of energy consumption per hour for 5 GUs in the same group.}
    \label{fig:5G1F_Energy}
\end{figure}

For the first scenario with 5 GUs, a maximum channel capacity of 250 Mbit/s was considered, since 5 GUs were sharing the same channel. The values of offered load were randomly defined. The results for energy consumption per hour are presented in Fig. \ref{fig:5G1F_Energy}. SUPPLY chose the Circular trajectory for the FAP leading to an energy consumption per hour of 457.58 kJ, implying an energy consumption reduction of 25\% in relation to the hovering state.

\begin{figure}[]
    \centering
    \includegraphics[width=0.45\textwidth]{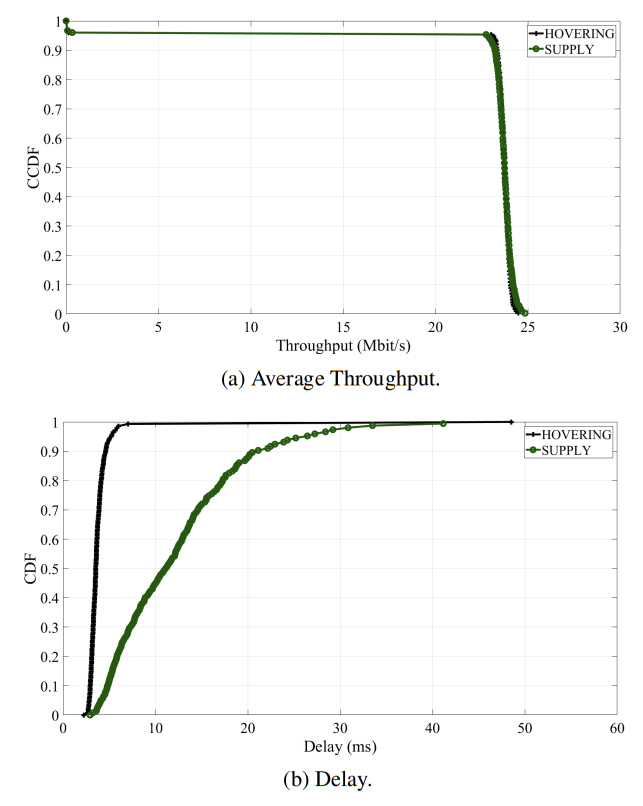}
    \caption{Network performance results for 5 GUs in the same group.}
    \label{fig:5G1F}
\end{figure}

In terms of network performance, the average throughput results show that there was no significant impact when using SUPPLY in relation to the hovering state ($<0.5\%$), allowing to guarantee the expected average throughput of 23.8 Mbit/s. When it comes to delay, there was an increase when following the SUPPLY-defined trajectory up to 20 ms for the $90^{th}$ percentile, as illustrated in Fig. \ref{fig:5G1F}. The increase in the delay is in line with the delay obtained for the networking scenario consisting of 2 GUs.

\begin{figure}[]
    \centering
    \includegraphics[width=0.45\textwidth]{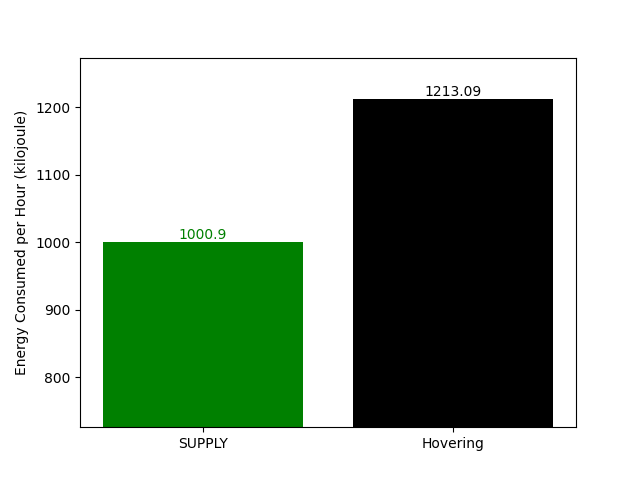}
    \caption{Comparison of energy consumption per hour for 5 GUs in 2 separate groups.}
    \label{fig:5G2F_Energy}
\end{figure}

The second scenario considered 5 GUs with randomly defined offered load values, considering a maximum channel capacity of 500 Mbit/s. The SUPPLY algorithm imposed the creation of 2 groups served by 2 FAPs. $FAP_{1}$ was serving a group composed of 3 GUs, while $FAP_{2}$ was serving a group composed of 2 GUs. The energy consumption per hour results can be seen in Fig. \ref{fig:5G2F_Energy}. In order to optimize energy consumption, the SUPPLY algorithm chose different trajectories for each FAP. For $FAP_{1}$, SUPPLY selected the Elliptic trajectory and for $FAP_{2}$ the Circular trajectory. This results in an energy consumption per hour of 1000.90 kJ, which is 18\% less when compared with 1213.09 kJ achieved for the hovering state.

\begin{figure}[]
    \centering
    \includegraphics[width=0.45\textwidth]{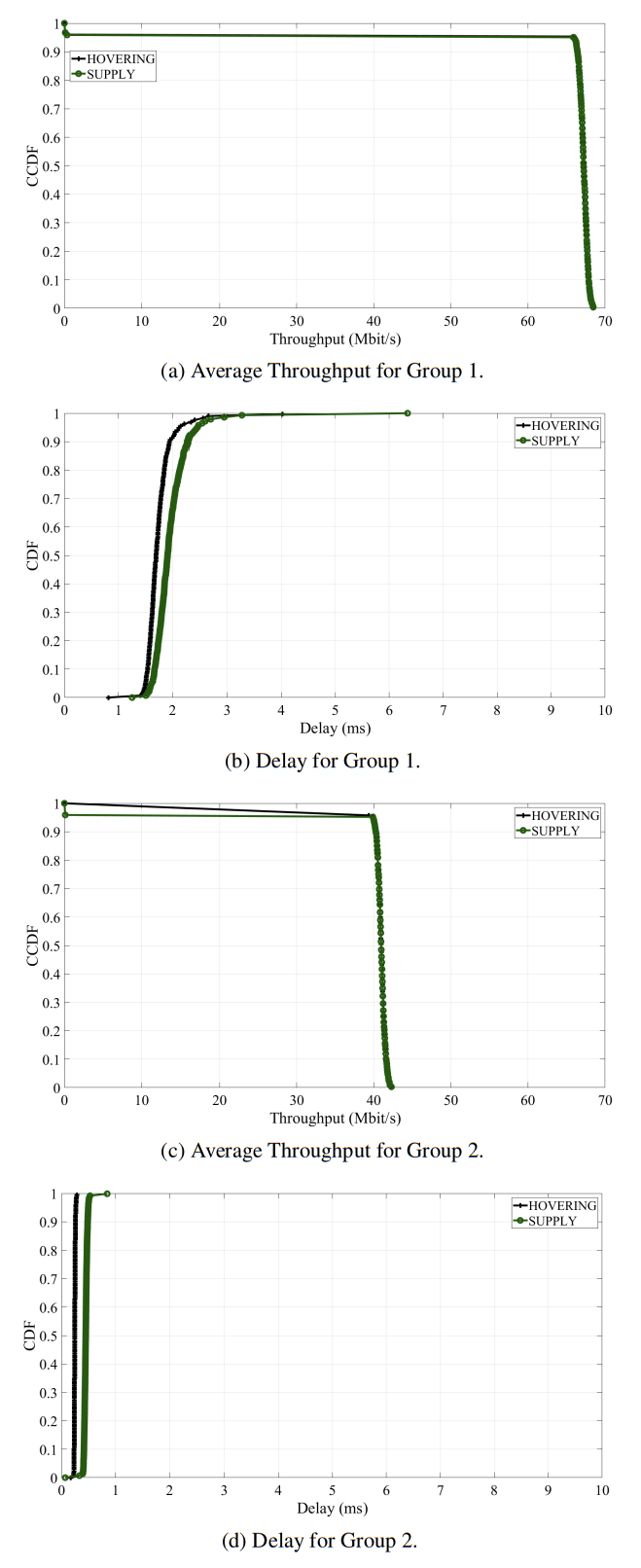}
    \caption{Network performance results for 5 GUs in separate groups.}
    \label{fig:5G2F}
\end{figure}

Regarding network performance, the average throughput results show that SUPPLY does not introduce a significant impact in relation to the hovering state, assuring 67.3 Mbit/s and 41.0 Mbit/s for group 1 and group 2, respectively, as illustrated in Fig. \ref{fig:5G2F}. No significant delay increase is observed, as shown in Fig. \ref{fig:5G2F}. 

\subsubsection{10 GUs}
\begin{figure}[]
    \centering
    \includegraphics[width=0.45\textwidth]{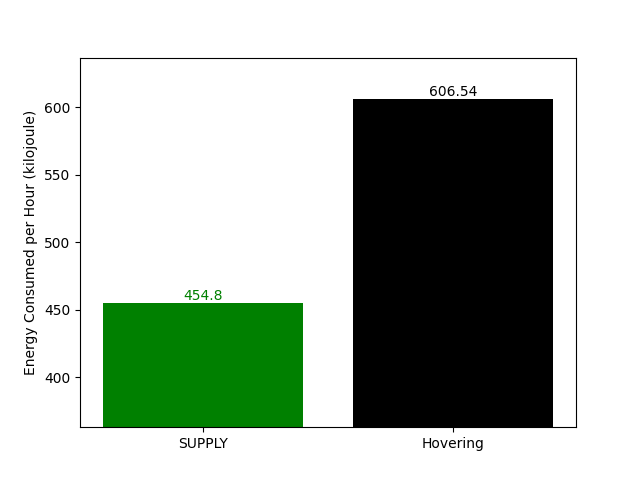}
    \caption{Comparison of energy consumption per hour for 10 GUs in the same group.}
    \label{fig:10G1F_Energy}
\end{figure}

The first scenario considered 10 GUs with randomly defined offered load values, while taking into account a maximum aggregated value of 100 Mbit/s since 10 GUs share the same channel. The results for the energy consumption per hour are presented in Fig. \ref{fig:10G1F_Energy}. The SUPPLY algorithm chose the Circular trajectory resulting in an energy consumption per hour of 454.80 kJ. This represents an energy consumption reduction of 25\% in relation to the hovering state.

\begin{figure}[]
    \centering
    \includegraphics[width=0.45\textwidth]{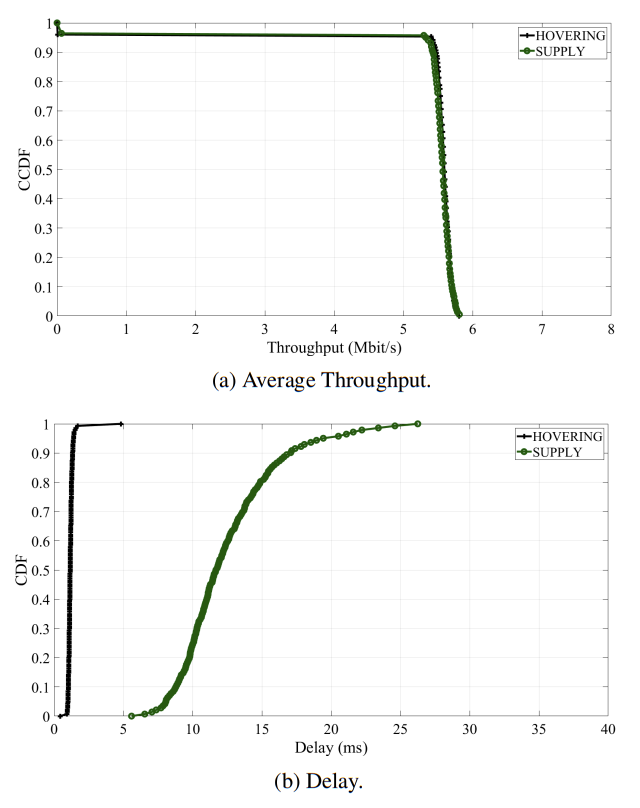}
    \caption{Network performance results for 10 GUs in the same group.}
    \label{fig:10G1F}
\end{figure}

In terms of network performance, the results show that the difference in average throughput between the SUPPLY and the hovering state is negligible, both achieving 5.6 Mbit/s. For delay, there is an increase when employing SUPPLY. The results show an increase between 5 and 15 ms, as depicted in Fig. \ref{fig:10G1F}. The increase in delay is justified by the greater distances between the FAP and the GUs that result from following the trajectory causing a decrease in SNR, as mentioned for the scenarios with 2 and 5 GUs.

\begin{figure}[]
    \centering
    \includegraphics[width=0.45\textwidth]{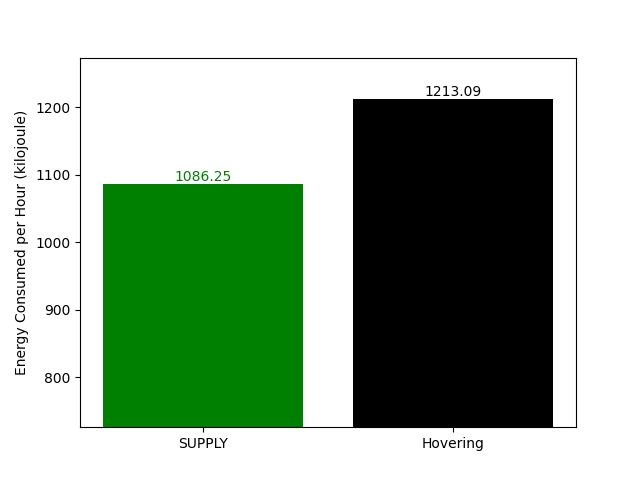}
    \caption{Comparison of energy consumption per hour for 10 GUs in separate groups.}
    \label{fig:10G2F_Energy}
\end{figure}

Finally, we considered a scenario with 10 GUs, for which the SUPPLY algorithm led to the creation of 2 groups served by 2 FAPs. One FAP serving 6 GUs and the other serving 4 GUs. In this scenario, the SUPPLY algorithm selected different trajectories for each FAP: $FAP_{1}$ followed the Elliptic trajectory and $FAP_{2}$ followed the Circular trajectory. SUPPLY led to an energy consumption per hour of 1086.25 kJ, while the hovering state resulted in 1213.09 kJ, as can be seen in Fig. \ref{fig:10G2F_Energy}. This represents a decrease of 11\% with respect to the hovering state. 

\begin{figure}[]
    \centering
    \includegraphics[width=0.45\textwidth]{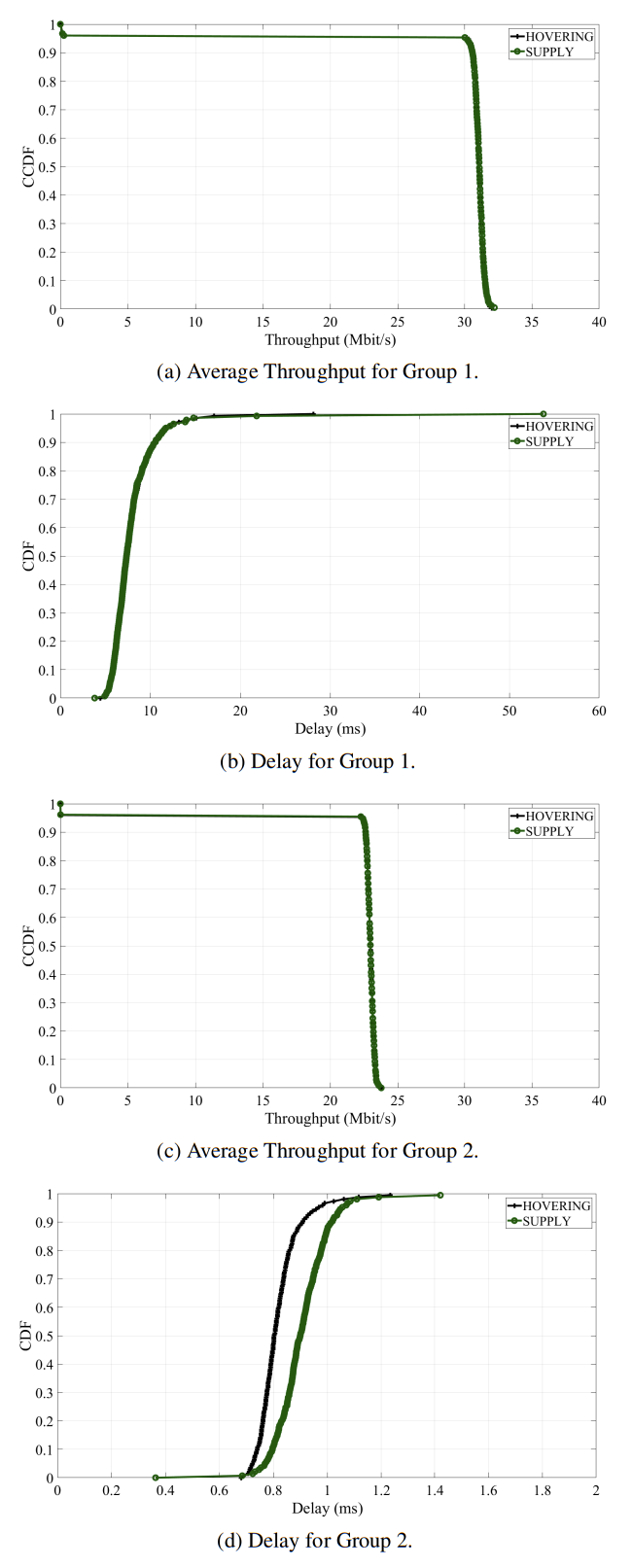}
    \caption{Network performance results for 10 GUs in separate groups.}
    \label{fig:10G2F}
\end{figure}

Regarding network performance, SUPPLY did not have a significant impact on the average throughput compared to the hovering state, assuring the expected averages of 31.2 Mbit/s and 23.0 Mbit/s, as shown in Fig. \ref{fig:10G2F}. No significant delay increases were observed (cf. Fig. \ref{fig:10G2F}).

\subsection{Energy Evaluation for Random Networking Scenarios}
\begin{figure}[]
    \centering
    \includegraphics[width=0.45\textwidth]{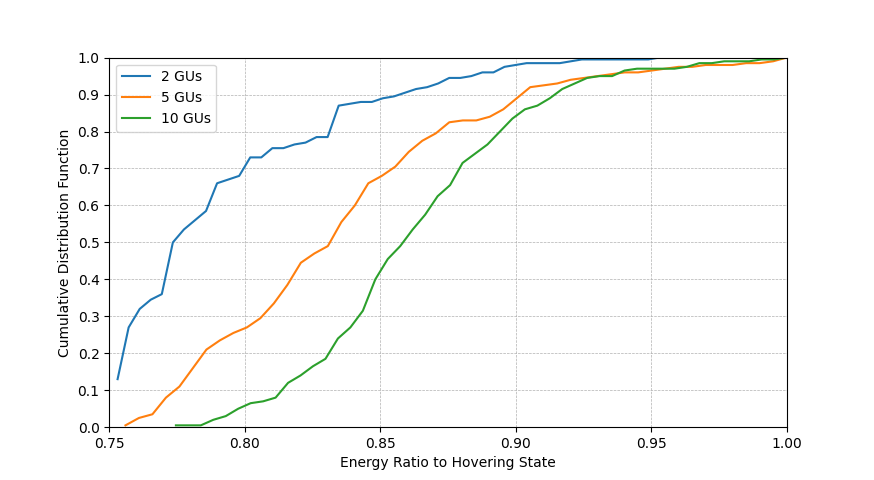}
    \caption{Energy ratios for the SUPPLY algorithm in relation to the hovering state, considering 200 random networking scenarios for each distinct number of GUs -- 2, 5, and 10.}
    \label{fig:CDF_200}
\end{figure}

Considering random networking scenarios with a given number of GUs and a predefined range of possible offered loads for the GUs, we evaluated how the number of GUs in the network influences the energy consumption.
This allowed assessing the gains achieved by the SUPPLY algorithm for generic networking scenarios.
For that purpose, we generated 200 random networking scenarios for different number of GUs in the network: 2, 5, and 10 GUs. The offered load values were randomly selected, considering a maximum channel capacity of 500 Mbit/s for all the scenarios. For evaluating the gains achieved by SUPPLY when it comes to energy consumption, we calculated the energy ratio in relation to the hovering state. The energy ratio metric consists of the SUPPLY energy consumption over the hovering state energy consumption. The results are shown in Fig. \ref{fig:CDF_200}. The $50^{th}$ percentile reveals energy ratios of 0.77, 0.83, and 0.86 for scenarios with 2, 5, and 10 GUs, respectively. On the other hand, the $90^{th}$ percentile shows energy ratios of 0.85, 0.90, and 0.91, respectively.

The obtained results allow us to conclude that as the number of GUs increases, the energy gains with respect to the hovering state decrease.
It is also possible to note an increase in the average number of FAPs needed with the increase in the number of GUs. For scenarios with 2, 5, and 10 GUs, the average number of FAPs is 1, 2, and 3, respectively.

\section{Discussion} \label{sec:Discussion}
From the results presented, it is possible to conclude that the SUPPLY algorithm results in reduced energy consumption with minimal impact on network performance.

Regarding the energy consumption results, increasing the number of GUs leads to less energy gains achieved by SUPPLY. Due to the increased number of GUs in the same area, the SUPPLY algorithm tends to generate larger groups that result in smaller intersection areas. With a smaller area to place the FAP, the SUPPLY algorithm defines less energy-efficient trajectories, since they are characterized by smaller straight line segments and radii. The intersection of the GUs spheres results in different shapes of intersection areas depending on the number of GUs in the group and their offered load. As such, different trajectories may be selected by SUPPLY for different shapes of intersection areas. It is observable that, for more Circular-shaped areas, the Circular trajectory is the preferable trajectory. On the other hand, for more elongated areas, the Elliptic trajectory leads to less energy consumption.

Regarding network performance, the SUPPLY trajectories in general allow achieving similar performance to the hovering state, even when considering a Rician fast-fading component in addition to the FSPL model.
The introduction of the Rician fast-fading component in scenarios where all the GUs share the same channel results in performance degradation, especially for cases where the FAPs complete the Circular and the Elliptic trajectories.
This degradation is expected since the Rician fast-fading component adds a random component to represent the variability observed in real-world wireless channels. The Circular and Elliptic trajectories are the most affected since they always reach the perimeter of the intersection area leading to higher SNR degradation.
However, the considered SNR margin of 1 dB in the SUPPLY algorithm allows to mitigate these losses.
The Inner Elliptic trajectory introduces a margin by not reaching the perimeter of the intersection areas in most cases. This has a similar effect to the addition of an SNR margin. However, since it does not take full advantage of the intersection area, the energy consumption results tend to be worse than the ones achieved by the other SUPPLY trajectories. For that reason, it is the least chosen trajectory by the SUPPLY algorithm.

Another aspect to take into account is the altitude at which the FAPs are positioned. For the evaluated scenarios we considered a 6 meter altitude, as mentioned in Section \ref{sec:SUPPLY}. However, due to factors, such as legislation and the FN application context, a higher altitude may need to be employed. Increasing the altitude leads to smaller intersection areas, thus resulting in less energy efficiency. Additionally, it may not be possible to assure the required QoS levels to the GUs, due to the increased distance between them and the FAPs. Finally, it is important to note that despite SUPPLY being evaluated for the IEEE 802.11ac Wi-Fi standard, it is thought to be applicable to other communications technologies.

\section{Conclusion} \label{sec:Conclusion}
UAVs have become increasingly popular in recent years due to their versatility. As UAVs are capable of hovering over the ground and carrying on-board cargo, they are suitable platforms for transporting communications nodes. For that reason, UAVs are an emerging solution to form FNs able to establish and reinforce wireless coverage and capacity on-demand. When compared with terrestrial networks, FNs bring up additional challenges. Since UAVs are not permanently connected to the power grid, they rely on on-board power sources, typically electric batteries, for both communications and movement.

We proposed SUPPLY, a new algorithm that enables the energy and performance-aware placement of multiple UAVs in an FN. SUPPLY takes into consideration the UAVs' energy consumption and the offered load of the GUs to define UAV trajectories that minimize the energy consumption while ensuring the targeted QoS requirements. The SUPPLY algorithm was evaluated in simulation in terms of energy consumption and network performance. The obtained results show that SUPPLY enables reduced energy consumption with minimal network performance impact.

As future work, an adaptive SNR margin can be implemented to further improve SUPPLY. Additionally, real-world experiments may be made to confirm the simulated energy consumption values and the network performance results. Finally, other trajectories may be investigated, and Machine Learning techniques may be used to learn trajectories that lead to higher energy efficiency gains. 

\bibliographystyle{IEEEtran}
\bibliography{refs}

\EOD

\end{document}